\documentclass[aps,reprint,prd,superscriptaddress,nofootinbib,amsmath,amssymb]{revtex4-2}

\usepackage{amssymb}
\usepackage[utf8]{inputenc}
\usepackage{amsmath}
\usepackage{amsfonts}
\usepackage{color}
\usepackage{graphicx}
\usepackage{verbatim}
\usepackage{mathrsfs}
\usepackage{hyperref}
\usepackage{orcidlink}

\def\rd{{\rm d}}

\def\rx{{\rm x}}
\def\ry{{\rm y}}

\begin{document}

\title{Ultrarelativistic limit of the Kerr theorem}

\author{Eloy \surname{Ay\'{o}n-Beato}\,\orcidlink{0000-0002-4498-3147}}
\email[]{eloy.ayon-beato@cinvestav.mx}
\affiliation{Departamento de F\'{i}sica, CINVESTAV-IPN, A.P. 14-740, C.P. 07360, Ciudad de M\'exico, Mexico}

\author{Daniel \surname{Flores-Alfonso}\,\orcidlink{0000-0001-7866-3531}}
\email[]{dafa@azc.uam.mx}
\affiliation{Departamento de Ciencias B\'asicas, Universidad Aut\'onoma Metropolitana -- Azcapotzalco, Avenida
San Pablo 420, Colonia Nueva El Rosario, Azcapotzalco 02128, Ciudad de M\'exico, Mexico}

\author{Mokhtar \surname{Hassaine}\,\orcidlink{0009-0003-3159-5916}}
\email[]{hassaine@inst-mat.utalca.cl}
\affiliation{Instituto de Matem\'{a}ticas (INSTMAT), Universidad de Talca, Casilla 747, Talca 3460000, Chile}

\begin{abstract}
The original Kerr theorem provides the foundation for Kerr–Schild transformations by classifying all shear-free and geodesic null congruences in flat spacetime; the key ingredient of the Kerr–Schild ansatz. However, due to the high level of degeneracy of the outcome it is often less practical than its symmetric refinements, which may single out congruences leading to physically significant spacetimes by imposing relevant symmetries. An illustrative example is the stationary axisymmetric version of Kerr theorem which has been shown to lead directly and uniquely to the Kerr black hole in vacuum. In this work, we propose a new symmetric refinement of the Kerr theorem by boosting the stationary symmetry into its ultrarelativistic limit to achieve invariance under null translations, while keeping axisymmetry. Under these assumptions, the classification yields only two distinct congruences. The first congruence is covariantly constant and, through the Kerr–Schild ansatz, evidently yields an axisymmetric \emph{pp}-wave. The vacuum axisymmetric profile of this \emph{pp}-wave displays a logarithmic dependence on the polar radius, characteristic of the exterior gravitational field of the  Bonnor light beam, and includes as a special case the Aichelburg–Sexl ultrarelativistic limit of the Schwarzschild black hole. The Kerr–Schild transformation of the second congruence gives rise to a non-trivial vacuum solution recently reported in [Phys.\ Rev.\ D \textbf{112}, 024020 (2025)]. Using circularity and appropriately fixing the reparameterization invariance of the orthogonal manifold to the Killing fields, we show that the latter solution corresponds to the well-known Taub–NUT spacetime with planar topology. These results emphasize how symmetry-based refinements of the Kerr theorem constitute a powerful tool to constructing physically essential spacetimes.
\end{abstract}

\maketitle

\section{Introduction\label{sec:Intro}}

Symmetries play a central role in theoretical physics. Beyond their ability to simplify the equations of motion or reveal conserved quantities via Noether’s theorem, symmetries often dictate which physical configurations are viable within a given theory. In classical mechanics, symmetries guide integrability and reduce degrees of freedom. In quantum field theory, internal and spacetime symmetries classify particles and govern interaction dynamics. In General Relativity, spacetime symmetries play a fundamental role in the study and classification of exact solutions to Einstein’s field equations. These symmetries not only simplify the equations of motion but also reveal essential physical properties of the gravitational field. For example, symmetries such as stationarity or axisymmetry typically correspond to conserved quantities like energy or angular momentum, providing deep insight into the physical interpretation of the spacetime beyond merely constraining the metric tensor. This makes symmetries indispensable tools for organizing and understanding exact solutions. A key framework  for ordering exact solutions arises from the algebraic classification of the Weyl tensor, originally introduced by Petrov~\cite{Petrov:1954,*Petrov:2000bs}. This scheme categorizes spacetimes according to the multiplicity and nature of the Weyl tensor eigenvalues or equivalently of their principal null directions, highlighting special algebraic structures. Among these, the algebraically special spacetimes, characterized by repeated principal null directions, are particularly significant because they often correspond to physically relevant solutions such as (rotating) black holes. On the other hand, the Goldberg–Sachs theorem provides a powerful geometric characterization of these algebraically special spacetimes~\cite{Goldberg:1962,*Goldberg:2009}. Concretely, the theorem states that a vacuum spacetime is algebraically special if and only if it admits a shear-free and geodesic null congruence. This theorem links the algebraic structure of the curvature with the existence of distinguished null congruences, which is both mathematically elegant and physically meaningful. Building upon these ideas, the Kerr theorem \cite{Debney:1969zz,Cox:1976,Stephani:2003tm} determines all shear-free and geodesic null congruences in flat spacetime. These congruences, through the Kerr–Schild ansatz \cite{Kerr:1965,*Kerr:1965vyg}, constitute a key ingredient in deriving nontrivial algebraically special solutions of Einstein’s equations. More precisely, in this construction the metric $g$ is represented as a linear ``perturbation'' of Minkowski metric $\eta$ built with the tangent vector $l$ to a shear-free and geodesic null congruence of flat spacetime,
\begin{equation}\label{eq:KSansatz}
g = \eta + 2S\, l \otimes l,
\end{equation}
where $S$ is a scalar profile. Remarkably, this ansatz linearizes the Einstein equations non-perturbatively in terms of $S$. Additionally, since the shear-free and geodesic null properties of the congruence are inherited in the new spacetime, the Kerr–Schild ansatz allows for the construction of nontrivial algebraically special solutions by focusing on the geometric properties of null congruences in flat spacetime. Thus, the Kerr theorem acts as a bridge that translates the problem of finding exact solutions into the study of kinematic properties of congruences, underlining the deep interplay between geometry, algebraic classification, and exact solutions in General Relativity.

However, the original formulation of the Kerr theorem suffers from a key ambiguity: it does not uniquely determine the congruence. In the absence of further constraints, there exists an infinite family of congruences satisfying the required properties, and thus a large degeneracy in the resulting metrics. This leads to a lack of uniqueness in the identification of the physically relevant solution, such as the Kerr black hole~\cite{Kerr:1963ud}. To overcome this limitation, a symmetry-enhanced refinement of the Kerr theorem has been proposed in~\cite{Ayon-Beato:2015nvz}. The essential idea is to impose physically relevant symmetries directly on the congruence, namely, stationarity and axisymmetry. These are precisely the symmetries expected of isolated rotating astrophysical objects and, in particular of the final state of gravitational collapse: the celebrated black holes. Under these constraints, the space of admissible congruences becomes drastically reduced, and one is left with a unique congruence compatible with both the geometric conditions and the physical symmetries. When substituted into the Kerr–Schild ansatz, this uniquely determines the Kerr metric~\cite{Kerr:1963ud} as the only solution consistent with the imposed symmetries. Moreover, in this framework, angular momentum naturally appears as the conserved charge associated with the axial symmetry of the congruence, and the physical relevance of the Kerr solution emerges directly from symmetry principles, without relying on arbitrary assumptions. This refinement illustrates the deep utility of symmetries not merely as computational tools, but as guiding physical principles that can constrain and determine the fundamental structure of spacetimes in General Relativity.

In this work, we adopt a strategy similar to that of \cite{Ayon-Beato:2015nvz}, but considering a different set of symmetries. Specifically, we propose a new symmetric refinement of the Kerr theorem where the usual assumption of stationarity is boosted into its ultrarelativistic regime and replaced by invariance under null translations. We thus impose both null translational invariance and axisymmetry, and classify all compatible shear-free and geodesic null congruences in flat spacetime. This classification reveals precisely two admissible congruences: one leading to axisymmetric \emph{pp}-waves, and the other enabling the construction of a nontrivial vacuum solution via the Kerr–Schild ansatz. Surprisingly, each congruence yields a unique vacuum solution of the Einstein equations. 

In the case of the vacuum axisymmetric \emph{pp}-wave, the associated profile exhibits a logarithmic dependence on the polar radius up-to the usual global scaling factor depending on the retarded time. This solution represents the exterior gravitational field of a cylindrical beam of light propagating along the axis of symmetry, with an intensity that varies along the retarded time \cite{Bonnor:1969mfs}. When this intensity is taken to be a Dirac delta distribution, the metric reduces to the renowned Aichelburg–Sexl geometry \cite{Aichelburg:1970dh}, which describes the ultrarelativistic limit of the Schwarzschild black hole, and has served as a cornerstone example in studies of high-energy gravitational interactions. 

For the second congruence, the Kerr–Schild ansatz leads to a distinct vacuum geometry, which coincides with the solution recently derived in Ref.~\cite{Harada:2025krm}. A closer analysis reveals that this metric corresponds to a planar version of the Taub–NUT spacetime. However, this identification is far from straightforward: the initial symmetric Kerr–Schild form hides the relation with the standard Taub–NUT structure. It is only through a detailed investigation of the circularity properties of the solution that the relevant diffeomorphism becomes apparent. The Taub–NUT solution itself holds a distinguished place among the exact solutions of General Relativity. Originally introduced by Taub~\cite{Taub:1950ez} and later extended by Newman, Tamburino, and Unti~\cite{Newman:1963yy}, it generalizes the Schwarzschild metric by including the so-called NUT parameter, often viewed as a gravitational analogue of a magnetic monopole. This geometry exhibits a number of fascinating and sometimes paradoxical features, such as closed timelike curves and the Misner string, a line singularity reminiscent of the Dirac string in gauge theory~\cite{Misner:1963fr}.

The remainder of the paper is organized as follows. In the next section, we present a new symmetric refinement of the Kerr theorem in the ultrarelativistic setting. This refinement focuses on shear-free and geodesic null congruences that are invariant under both advanced null translations and axisymmetry. The classification yields two distinct congruences: one covariantly constant, and the other is precisely the one used to build the geometry studied in Ref.~\cite{Harada:2025krm}. In Sec.~\ref{sec:KS}, we analyze the latter congruence in detail. We show that a direct application of the circularity theorem for vacuum Einstein solutions constrains the coordinate dependence of the profile function $S$ in the Kerr–Schild ansatz \eqref{eq:KSansatz}. Substituting this ansatz into the vacuum field equations reproduces the exact metric proposed in Ref.~\cite{Harada:2025krm}. After exploiting the geometric consequences of the proved circularity, we then demonstrate that this metric is in fact the Taub–NUT spacetime with planar base manifold, by performing an explicit coordinate transformation that brings it into its standard form. This procedure allows us to identify the mass and NUT parameters, as well as the underlying planar topology. In Sec.~\ref{sec:PP} we turn to the covariantly constant congruence, consequently corresponding to axisymmetric \emph{pp}-waves via the Kerr–Schild ansatz. In this case, the circularity condition is automatically satisfied, and the vacuum equations uniquely fix the profile function to be proportional to the logarithm of the polar radius. We discuss the properties of the resulting \emph{pp}-wave metric, which represents the unique axisymmetric Kerr–Schild vacuum solution within this class. The final section summarizes our results and offers concluding remarks.

\section{A new symmetric Kerr theorem\label{sec:KT}}

Here, we consider the ultrarelativistic limit of the stationary axisymmetric Kerr theorem proven in~\cite{Ayon-Beato:2015nvz}. Specifically, we focus on characterizing  shear-free and geodesic null congruences in flat spacetime 
\begin{equation}\label{eq:flat}
\rd s^2 = -2\rd u \rd v + \rd \rho^2 + \rho^2 \rd\phi^2,
\end{equation}
which are invariant under advanced null translations and axisymmetry, described by the Killing fields $k = \partial_v$ and $m = \partial_\phi$, respectively. We closely follow the successful strategy proposed in~\cite{Ayon-Beato:2015nvz}. The starting point is to note that the tangent vector to any congruence compatible with these symmetries must necessarily takes the form
\begin{equation}\label{eq:l_mu_null_axi}
l = l_v(u,\rho) \rd v + l_u(u,\rho) \rd u + l_\rho(u,\rho) \rd \rho + l_\phi(u,\rho) \rd \phi.
\end{equation}
The geodesic equation for such vectors, $l^{\alpha} \nabla_{\alpha} l_{\beta} = 0$, becomes
\begin{equation}\label{eq:GeoCl}
(l_\rho \partial_\rho - l_v \partial_u) l_{\beta} =
\frac{l_\phi^2}{\rho^3} \delta_{\beta}^\rho.
\end{equation}
The equation for component $l_\rho$ is automatically satisfied thanks to the null condition if the others hold, since $l^\mu \nabla_\mu (l_{\nu} l^{\nu}) = 2 l^{\nu} l^\mu \nabla_{\mu} l_{\nu}$. The remaining geodesic equations imply the other components are invariant along the integral curves of the contravariant version of the vector field
\begin{equation}\label{eq:l^mu}
l = -l_u \partial_v - l_v \partial_u + l_\rho \partial_\rho + \frac{l_\phi}{\rho^2} \partial_\phi,
\end{equation}
which is compactly expressed as $l(l_v) = l(l_u) = l(l_\phi) = 0$. This is consistent with the fact that these quantities correspond to the conserved momenta  conjugate to the cyclic coordinates of the geodesic Lagrangian. All the involved homogeneous linear first-order PDE share the same bidimensional characteristic system
\begin{equation}\label{eq:char}
\frac{\rd \rho}{l_\rho} = -\frac{\rd u}{l_v},
\end{equation}
which allows for a single independent invariant. Consequently, all other invariants must be functions of this one, and we take $l_u$ to play this role. Furthermore, for $l_v \ne 0$, we can use the scaling freedom in the affine parameterization to fix $l_v = 1$. Observe that in the degenerate case $l_v = 0$, the null condition implies $l_\rho = 0 = l_\phi$, and the resulting tangent vector reduces to $l = \rd u$, which is covariantly constant and trivially satisfies all required conditions.

Recapitulating the geodesic characterization, the tangent vector to the most general geodesic null congruence in flat spacetime invariant under advanced null translations and axisymmetry is given, in the generic case $l_v \ne 0$, by
\begin{subequations}\label{eq:NullGCl}
\begin{equation}
l = \rd v + l_u \rd u + l_\rho \rd \rho + l_\phi(l_u) \rd \phi,
\end{equation}
where the component $l_\rho = l_\rho(u,\rho)$ is determined from the null condition
\begin{equation}\label{eq:NullCond}
l_\rho^2 = 2l_u - \frac{l_\phi(l_u)^2}{\rho^2},
\end{equation}
and the component $l_u = l_u(u,\rho)$ is an invariant of the geodesic motion, namely
\begin{equation}\label{eq:l(l_u)=0}
l(l_u) = (l_\rho \partial_\rho - l_u \partial_u) l_u = 0.
\end{equation}
\end{subequations}

The shear-free condition is equivalent to demanding the vanishing of the following quadratic quantity, see~\cite{Ayon-Beato:2015nvz},
\begin{equation}\label{eq:Sigma2}
2\sigma^2\equiv\sigma_{\mu\nu}\sigma^{\mu\nu}
=\nabla_{(\mu}l_{\nu)}\nabla^{\mu}l^{\nu}
-\frac{1}{2}\left(\nabla_{\mu}l^{\mu}\right)^2,
\end{equation}
which for the geodesic null congruence \eqref{eq:NullGCl} becomes
\begin{align}
2\sigma^2={}&\frac{l_u}{\rho^2} 
\Biggl[\frac{( \rho\partial_\rho l_u - 2 l_u )^2}{4l_u^2} \nonumber \\
&+ \frac{1}{l_\rho^2} \left(
\partial_\rho l_u\frac{\rd l_\phi}{\rd l_u}
- \frac{l_\phi( \rho\partial_\rho l_u + 2 l_u )}{2\rho l_u}
\right)^2\Biggr]=0, \label{eq:shear}
\end{align}
after using the geodesic equation \eqref{eq:l(l_u)=0}. Inserting the condition that follows from the first square above in that one which follows from the second, we obtain
\begin{equation}\label{eq:ODE(l[phi])}
\frac{\rd l_\phi\left(l_u\right)}{\rd l_u}=\frac{l_\phi(l_u)}{l_u},
\end{equation}
which straightforwardly integrates to
\begin{equation}\label{eq:l_phi}
l_\phi\left(l_u\right)=2 b l_u,
\end{equation}
where the integration constant $b$ tempers the angular-momentum conservation of the congruence that follows from axisymmetry. We now turn to the only two remaining constraints, the first shear-free condition in \eqref{eq:shear} and the geodesic equation \eqref{eq:l(l_u)=0}, both define the system
\begin{align}\label{eq:l_u(u,rho)}
\partial_\rho l_u &= \frac{2 l_u}{\rho}, & \partial_u l_u &= \frac{2 l_u l_\rho}{\rho},
\end{align}
which is integrable, as the mixed partial derivatives commute $\partial_u\partial_\rho l_u = \partial_\rho\partial_u l_u$. The first equation easily integrates as
\begin{equation}\label{eq:f(z)def}
l_u = \rho^2 f(z),
\end{equation}
and, using the null condition \eqref{eq:NullCond}, the second equation reduces to the ordinary differential equation
\begin{equation}\label{eq:f(z)}
\left( \sqrt{\frac{1}{2f} - b^2} \right)' = -1, \quad \Longrightarrow \quad
\sqrt{\frac{1}{2f} - b^2} = -(u - u_0).
\end{equation}
Exploiting the additional retarded-null isometry of flat spacetime generated by the Killing vector $\partial_u$, we can set $u_0 = 0$ without loss of generality. This concludes the construction.

In summary, we have established the following ultrarelativistic version of the Kerr theorem: The tangent vectors to the most general shear-free and geodesic null congruences in flat spacetime that are invariant under advanced null translations and axisymmetry fall into exactly two classes, given by
\begin{subequations}
\begin{align}
l &= \rd u,\label{eq:lcc}\\
l &= \rd v+\frac{\rho^2\rd u}{2(u^2 + b^2)}  - \frac{u\rho\rd\rho}{u^2 + b^2} + \frac{b\rho^2\rd \phi}{u^2 + b^2}.\label{eq:l_Harada}
\end{align}
\end{subequations}
The congruence associated with the first vector is trivial, since it is covariantly constant. Remarkably, the second congruence coincides with the one recently introduced in \cite{Harada:2025krm}, where it was motivated from the Hopf fibration.

It follows from this result that any Kerr–Schild ansatz \eqref{eq:KSansatz} invariant under advanced null translations and axisymmetry must be built from one of the two congruences above. In the next two sections, we separately analyze the corresponding vacuum solutions, and demonstrate the uniqueness of the Kerr–Schild transformations with such symmetries within each class.

\section{Uniqueness of the Kerr–Schild vacuum 
with Hopf structure\label{sec:KS}}

Let us first consider the Kerr–Schild ansatz invariant under advanced null translations and axisymmetry associated to the nontrivial congruence \eqref{eq:l_Harada}
\begin{align}\label{eq:symmKS}
\rd s^2 ={}& -2\rd u \rd v +\rd \rho^2 + \rho^2\rd\phi^2+2S(u,\rho)\nonumber \\
& \times\left(\rd v+\frac{\rho^2\rd u}{2(u^2 + b^2)}  - \frac{u\rho\rd\rho}{u^2 + b^2} + \frac{b\rho^2\rd \phi}{u^2 + b^2} \right)^2.
\end{align}
We now demonstrate that there exists a unique vacuum solution within this class.

The starting point is a version of the circularity theorem, typically studied in the context of stationary axisymmetric spacetimes~\cite{Heusler:1996}, which remains valid under the present symmetry assumptions. Any pair of commuting Killing vectors, such as $k=\partial_v$ and $m=\partial_\phi$, satisfies the identities
\begin{equation}\label{eq:Circularity}
\begin{aligned}
    \rd \star (k \wedge m \wedge \rd k) &= 2\star\left(k \wedge m \wedge R(k)\right),\\
    \rd \star (k \wedge m \wedge \rd m) &= 2\star\left(k \wedge m \wedge R(m)\right),
\end{aligned}
\end{equation}
where $k$ and $m$ are understood as 1-forms, and $R(k) = R_{\mu\nu}k^\nu \rd x^\mu$, $R(m) = R_{\mu\nu}m^\nu \rd x^\mu$ denote the corresponding Ricci 1-forms. In vacuum, the smooth functions under differentiation at the left-hand sides become globally defined constants, which in turn vanish on the symmetry axis where $m = 0$. This leads to the Frobenius integrability conditions, which for the Kerr–Schild ansatz \eqref{eq:symmKS} reduce to 
\begin{equation}\label{eq:Frobenius}
\begin{alignedat}{2}
    0&=\star (k \wedge m \wedge \rd k) &&= 2\rho\partial_\rho S(u,\rho),\\
    0&=\star (k \wedge m \wedge \rd m) &&= \frac{2b\rho^3}{u^2+b^2}\partial_\rho S(u,\rho),
\end{alignedat}
\end{equation}
showing that the scalar profile $S(u,\rho)$ must be independent of the polar radius  $\rho$, i.e.\ $S(u,\rho) = S(u)$. This constraint is the starting point of \cite{Harada:2025krm}, where it was shown that the Einstein equations fully determine this special profile. The resulting unique Kerr–Schild vacuum solution for the studied congruence is
\begin{align}\label{eq:Harada}
\rd s^2 ={}& -2\rd u \rd v +\rd \rho^2 + \rho^2\rd\phi^2+\frac{N u}{u^2+b^2}\nonumber \\
& \times\left(\rd v+\frac{\rho^2\rd u}{2(u^2 + b^2)}  - \frac{u\rho\rd\rho}{u^2 + b^2} + \frac{b\rho^2\rd \phi}{u^2 + b^2} \right)^2.
\end{align}

Now recall that the Frobenius conditions \eqref{eq:Frobenius} ensure the integrability of the distribution orthogonal to the Killing vector fields, allowing the local definition of surfaces orthogonal to the symmetry directions to be extended globally. Introducing coordinates adapted to this foliation leads to a Boyer–Lindquist-type chart~\cite{Boyer:1966qh}, in which the metric takes a block-diagonal form. However, the Kerr–Schild representation \eqref{eq:Harada} additionally involves mixing between the non-Killing directions $(\rd u, \rd \rho)$. To bring this two-dimensional sector into the standard orthogonal form, we supplement the Boyer–Lindquist transformation with an additional reparameterization of this orthogonal manifold. The resulting coordinate transformation is
\begin{subequations}
\begin{align}
    v&=-t+\frac{r^2+2b^2\ln(r)}{2N} + \frac{r\varrho^2}{2},\\ 
    u&=r,\\
    \rho&= \varrho\sqrt{r^2+b^2},\\
    \phi&= \varphi-\arctan\left(\frac{r}{b}\right),
\end{align}
\end{subequations}
and the metric \eqref{eq:Harada} now takes the form
\begin{align}
    \rd s^2 = &-\left(\frac{-Nr}{r^2+b^2}\right)(\rd t -b\varrho^2\rd\varphi)^2+\left(\frac{r^2+b^2}{-Nr}\right)\rd r^2 \notag\\
    &+\left(r^2+b^2\right)\left(\rd\varrho^2 + \varrho^2\rd\varphi^2\right),
    \label{eq:Taub–NUT}
\end{align}
which is precisely the well-known planar Taub–NUT spacetime~\cite{Taub:1950ez,Newman:1963yy}. In this form the parameter $b$ is recognized as the NUT parameter and the integration constant $N$ as twice the usual mass parameter. Moreover, the complete isometry algebra of the metric is now discernible. By simply pushing forward the Killing vectors of the Taub–NUT metric \eqref{eq:Taub–NUT} one finds that metric \eqref{eq:Harada} admits two additional Killing vector fields to $k$ and $m$ given by
\begin{subequations}
    \begin{align}
   \xi_1 &= \rho\cos\phi\partial_v+\rx\partial_\rho+\frac{\ry}{\rho}\partial_\phi
   ,\\
    \xi_2 &= \rho\sin\phi\partial_v-\ry\partial_\rho+\frac{\rx}{\rho}\partial_\phi,
\end{align}
\end{subequations}
where $\rx = u\cos\phi+b\sin\phi$ and $\ry=-u\sin\phi+b\cos\phi$.

\section{Uniqueness of the axisymmetric \emph{pp}-wave vacuum\label{sec:PP}}

It follows from Sec.~\ref{sec:KT} that the Kerr–Schild ansatz invariant under advanced null translations and axisymmetry associated with the covariantly constant congruence \eqref{eq:lcc} can be rewritten as
\begin{equation}\label{eq:pp-wave}
\rd s^2 = - F(u,\rho)\rd u^2 -2\rd u \rd v + \rd \rho^2 + \rho^2 \rd\phi^2.
\end{equation}
This metric represents a family of axisymmetric plane-fronted gravitational waves with parallel rays (\emph{pp}-waves). Here, contrary to the previous section, the circularity imposes no restriction on the axisymmetric profile $F(u,\rho)$. When imposing vacuum Einstein equations for the axisymmetric \emph{pp}-waves, the profile satisfies an Euler-type differential equation in the polar radius $\rho$, characterized by a degenerate characteristic polynomial. The general solution takes then the following form, after a supertranslation of the coordinate $v$ \cite{Stephani:2003tm},
\begin{equation}\label{eq:pp-wave_sol}
F(u,\rho) = f(u)\ln \rho,
\end{equation}
where $f(u)$ is an arbitrary function of the retarded time $u$. The spacetime described by metric \eqref{eq:pp-wave} with profile \eqref{eq:pp-wave_sol} is not new, it corresponds to the exterior gravitational field of the Bonnor light beam \cite{Bonnor:1969mfs}. It is a classical exact solution representing an axisymmetric \emph{pp}-wave spacetime sourced by a straight beam of pure radiation traveling along the symmetry axis, where the corresponding interior solution is an appropriately glued plane wave. 

A particularly important and physically insightful special case arises when the beam is infinitely
thin and the profile function $f(u)$ is chosen to be proportional to a Dirac delta distribution in the retarded time $u$, i.e., $f(u) = \alpha\, \delta(u)$ for some constant amplitude $\alpha$. This choice yields an impulsive gravitational wave localized precisely on the null hypersurface $u = 0$, with the gravitational field sharply concentrated along this wavefront and vanishing elsewhere. In this impulsive limit, the resulting geometry describes the gravitational field generated by a massless particle moving at the speed of light, known as the Aichelburg–Sexl solution \cite{Aichelburg:1970dh}. This solution can be derived as the ultrarelativistic limit of the Schwarzschild metric when subjected to an infinite Lorentz boost in the direction of motion. The Aichelburg–Sexl spacetime has been fundamental in the study of gravitational shockwaves and high-energy collisions in General Relativity.

Thus, this analysis demonstrates that the analyzed Kerr–Schild ansatz not only encompasses the general structure of axisymmetric \emph{pp}-wave spacetimes but also naturally includes physically important limiting cases corresponding to ultrarelativistic boosts of well-known black hole solutions.

\section{Conclusions }

The motivation for this work stems from the recent solution presented in~\cite{Harada:2025krm}, which was not initially recognized as a form of the Taub–NUT spacetime with trivial topology. Its construction relies on a specific null congruence whose origin was unclear. This prompted us to investigate whether such a congruence could emerge naturally from a symmetry principle. To this end, we have developed a symmetric refinement of the Kerr theorem applicable in the ultrarelativistic regime, where invariance under null translations replaces the usual notion of stationarity. This refinement drastically reduces the degeneracy of the original classification and selects only two distinct shear-free and geodesic null congruences in flat spacetime. Our results show that symmetry constraints can effectively isolate physically relevant congruences, thereby guiding the construction of exact solutions to Einstein’s equations. Much like the stationary and axisymmetric refinement that yields the Kerr solution~\cite{Ayon-Beato:2015nvz}, our approach illustrates the power of symmetry-based selection criteria in identifying meaningful spacetimes within broader families of solutions.

In our search for invariant congruences, we uncover a notable feature of the integration constant $b$ in \eqref{eq:l_phi}, later identified as the NUT parameter. This constant can be seen as governing the conservation of angular momentum in geodesic motion, a property closely tied to axisymmetry, much like the angular momentum parameter $a$ in the Kerr solution derivation via the stationary and axisymmetric refinement of the Kerr theorem \cite{Ayon-Beato:2015nvz}. This interpretation is consistent with the fact that the Taub–NUT geometry exhibits nonvanishing angular momentum when the topology is trivial~\cite{Astefanesei:2004kn}. It thus offers deeper insight into the connection between geometric parameters and conserved quantities in these refined Kerr–Schild spacetimes.

Also, once these congruences are determined, extending the vacuum solutions to include Maxwell fields becomes a straightforward procedure. Indeed, by choosing the electromagnetic vector potential proportional to the null congruence itself, one can construct exact charged solutions within the Kerr–Schild framework \cite{Ayon-Beato:2015nvz}. This approach naturally generalizes both the axisymmetric \emph{pp}-wave and the planar Taub–NUT  solutions to their charged counterparts, preserving the underlying symmetries and geometric structure. In this sense, the symmetry-guided identification of congruences not only provides clarity on the vacuum solutions but also offers a systematic pathway to charged extensions, enriching the comprehension of physically relevant exact solutions in General Relativity. 

As made evident in our derivation, the shear-free nature of the congruence plays a crucial role in effectively selecting a (nearly) unique geodesic and null congruence. It is well known that in higher dimensions, the congruences underlying solutions such as the Myers–Perry black holes \cite{Myers:1986un} have generally nontrivial shear. Indeed, requiring the shear matrix to vanish identically, as in four dimensions, becomes overly restrictive as the number of dimensions increases. Nevertheless, it would be interesting to investigate whether weaker geometric constraints on the shear matrix could still suffice to uniquely characterize the congruences suited for a successful Kerr–Schild ansatz in higher dimensions. Understanding these aspects may prove valuable for extending the Kerr theorem to higher dimensions, where the ability to judiciously select congruences could lead to novel and physically meaningful solutions. In this sense, the framework developed in the present work may serve as a foundation for such generalizations.

Finally, it is both natural and desirable that physically relevant exact solutions in General Relativity and beyond, particularly those involving matter fields, be constructible from clear and robust physical principles rather than solely by ``chance'' or brute force. The success of the original Kerr theorem, in its stationary and axisymmetric formulation is a significative example. Indeed, by exploiting symmetries and the geometry of null congruences, it uniquely characterizes the Kerr black hole solution, a cornerstone of classical and astrophysical black hole physics. Similarly, the present work’s identification of the planar Taub–NUT solution within a symmetry-enhanced Kerr–Schild framework illustrates how imposing physically meaningful symmetry constraints can isolate and clarify important, albeit less conventional, exact spacetimes. Extending these ideas to theories with sources, such as scalar fields, or more general matter content, remains a crucial and largely open challenge. Developing symmetry-based selection criteria and geometric characterizations of null congruences in such contexts could provide systematic pathways to discovering new exact solutions that are both mathematically elegant and physically insightful. This approach would not only deepen our understanding of the interplay between geometry, symmetry, and matter in the presence of gravity but also guide the search for novel spacetimes relevant to astrophysics, cosmology, and fundamental physics.

\begin{acknowledgments}
We thank Adolfo Cisterna and Borja Diez for bringing this issue to our attention and Pedro A. S\'anchez for helpful discussions. D.F.-A.\ acknowledges financial support from SECIHTI through a postdoctoral research grant. M.H.\ gratefully acknowledges the University of Paris-Saclay for its warm hospitality during the development of this project.
\end{acknowledgments}


\begin{thebibliography}{21}%
\makeatletter
\providecommand \@ifxundefined [1]{%
 \@ifx{#1\undefined}
}%
\providecommand \@ifnum [1]{%
 \ifnum #1\expandafter \@firstoftwo
 \else \expandafter \@secondoftwo
 \fi
}%
\providecommand \@ifx [1]{%
 \ifx #1\expandafter \@firstoftwo
 \else \expandafter \@secondoftwo
 \fi
}%
\providecommand \natexlab [1]{#1}%
\providecommand \enquote  [1]{``#1''}%
\providecommand \bibnamefont  [1]{#1}%
\providecommand \bibfnamefont [1]{#1}%
\providecommand \citenamefont [1]{#1}%
\providecommand \href@noop [0]{\@secondoftwo}%
\providecommand \href [0]{\begingroup \@sanitize@url \@href}%
\providecommand \@href[1]{\@@startlink{#1}\@@href}%
\providecommand \@@href[1]{\endgroup#1\@@endlink}%
\providecommand \@sanitize@url [0]{\catcode `\\12\catcode `\$12\catcode `\&12\catcode `\#12\catcode `\^12\catcode `\_12\catcode `\%12\relax}%
\providecommand \@@startlink[1]{}%
\providecommand \@@endlink[0]{}%
\providecommand \url  [0]{\begingroup\@sanitize@url \@url }%
\providecommand \@url [1]{\endgroup\@href {#1}{\urlprefix }}%
\providecommand \urlprefix  [0]{URL }%
\providecommand \Eprint [0]{\href }%
\providecommand \doibase [0]{https://doi.org/}%
\providecommand \selectlanguage [0]{\@gobble}%
\providecommand \bibinfo  [0]{\@secondoftwo}%
\providecommand \bibfield  [0]{\@secondoftwo}%
\providecommand \translation [1]{[#1]}%
\providecommand \BibitemOpen [0]{}%
\providecommand \bibitemStop [0]{}%
\providecommand \bibitemNoStop [0]{.\EOS\space}%
\providecommand \EOS [0]{\spacefactor3000\relax}%
\providecommand \BibitemShut  [1]{\csname bibitem#1\endcsname}%
\let\auto@bib@innerbib\@empty
\bibitem [{\citenamefont {Petrov}(1954)}]{Petrov:1954}%
  \BibitemOpen
  \bibfield  {author} {\bibinfo {author} {\bibfnamefont {A.~Z.}\ \bibnamefont {Petrov}},\ }\href@noop {} {\bibfield  {journal} {\bibinfo  {journal} {Uch. Zapiski Kazan. Gos. Univ.}\ }\textbf {\bibinfo {volume} {114}},\ \bibinfo {pages} {55} (\bibinfo {year} {1954})}\BibitemShut {NoStop}%
\bibitem [{\citenamefont {Petrov}(2000)}]{Petrov:2000bs}%
  \BibitemOpen
  \bibfield  {author} {\bibinfo {author} {\bibfnamefont {A.~Z.}\ \bibnamefont {Petrov}},\ }\href {https://doi.org/10.1023/A:1001910908054} {\bibfield  {journal} {\bibinfo  {journal} {Gen. Rel. Grav.}\ }\textbf {\bibinfo {volume} {32}},\ \bibinfo {pages} {1661} (\bibinfo {year} {2000})}\BibitemShut {NoStop}%
\bibitem [{\citenamefont {Goldberg}\ and\ \citenamefont {Sachs}(1962)}]{Goldberg:1962}%
  \BibitemOpen
  \bibfield  {author} {\bibinfo {author} {\bibfnamefont {J.~N.}\ \bibnamefont {Goldberg}}\ and\ \bibinfo {author} {\bibfnamefont {R.~K.}\ \bibnamefont {Sachs}},\ }\href@noop {} {\bibfield  {journal} {\bibinfo  {journal} {Acta Phys. Pol.}\ }\textbf {\bibinfo {volume} {22}},\ \bibinfo {pages} {13} (\bibinfo {year} {1962})}\BibitemShut {NoStop}%
\bibitem [{\citenamefont {Goldberg}\ and\ \citenamefont {Sachs}(2009)}]{Goldberg:2009}%
  \BibitemOpen
  \bibfield  {author} {\bibinfo {author} {\bibfnamefont {J.~N.}\ \bibnamefont {Goldberg}}\ and\ \bibinfo {author} {\bibfnamefont {R.~K.}\ \bibnamefont {Sachs}},\ }\href {https://doi.org/10.1007/s10714-008-0722-5} {\bibfield  {journal} {\bibinfo  {journal} {Gen. Rel. Grav.}\ }\textbf {\bibinfo {volume} {41}},\ \bibinfo {pages} {433–} (\bibinfo {year} {2009})}\BibitemShut {NoStop}%
\bibitem [{\citenamefont {Debney}\ \emph {et~al.}(1969)\citenamefont {Debney}, \citenamefont {Kerr},\ and\ \citenamefont {Schild}}]{Debney:1969zz}%
  \BibitemOpen
  \bibfield  {author} {\bibinfo {author} {\bibfnamefont {G.~C.}\ \bibnamefont {Debney}}, \bibinfo {author} {\bibfnamefont {R.~P.}\ \bibnamefont {Kerr}},\ and\ \bibinfo {author} {\bibfnamefont {A.}~\bibnamefont {Schild}},\ }\href {https://doi.org/10.1063/1.1664769} {\bibfield  {journal} {\bibinfo  {journal} {J. Math. Phys.}\ }\textbf {\bibinfo {volume} {10}},\ \bibinfo {pages} {1842} (\bibinfo {year} {1969})}\BibitemShut {NoStop}%
\bibitem [{\citenamefont {Cox}\ and\ \citenamefont {Flaherty}(1976)}]{Cox:1976}%
  \BibitemOpen
  \bibfield  {author} {\bibinfo {author} {\bibfnamefont {D.}~\bibnamefont {Cox}}\ and\ \bibinfo {author} {\bibfnamefont {E.~J.}\ \bibnamefont {Flaherty}},\ }\href {https://doi.org/10.1007/BF01609355} {\bibfield  {journal} {\bibinfo  {journal} {Commun. Math. Phys.}\ }\textbf {\bibinfo {volume} {47}},\ \bibinfo {pages} {75} (\bibinfo {year} {1976})}\BibitemShut {NoStop}%
\bibitem [{\citenamefont {Stephani}\ \emph {et~al.}(2003)\citenamefont {Stephani}, \citenamefont {Kramer}, \citenamefont {MacCallum}, \citenamefont {Hoenselaers},\ and\ \citenamefont {Herlt}}]{Stephani:2003tm}%
  \BibitemOpen
  \bibfield  {author} {\bibinfo {author} {\bibfnamefont {H.}~\bibnamefont {Stephani}}, \bibinfo {author} {\bibfnamefont {D.}~\bibnamefont {Kramer}}, \bibinfo {author} {\bibfnamefont {M.~A.~H.}\ \bibnamefont {MacCallum}}, \bibinfo {author} {\bibfnamefont {C.}~\bibnamefont {Hoenselaers}},\ and\ \bibinfo {author} {\bibfnamefont {E.}~\bibnamefont {Herlt}},\ }\href {https://doi.org/10.1017/CBO9780511535185} {\emph {\bibinfo {title} {{Exact solutions of Einstein's field equations}}}}\ (\bibinfo  {publisher} {Cambridge Univ. Press},\ \bibinfo {year} {2003})\BibitemShut {NoStop}%
\bibitem [{\citenamefont {Kerr}\ and\ \citenamefont {Schild}(1965)}]{Kerr:1965}%
  \BibitemOpen
  \bibfield  {author} {\bibinfo {author} {\bibfnamefont {R.~P.}\ \bibnamefont {Kerr}}\ and\ \bibinfo {author} {\bibfnamefont {A.}~\bibnamefont {Schild}},\ }in\ \href@noop {} {\emph {\bibinfo {booktitle} {{Atti del Convegno sulla Relativit{\`{a}} Generale: Problemi dell'Energia e Onde Gravitazionali}}}}\ (\bibinfo  {publisher} {G. Barb{\`{e}}ra Editore},\ \bibinfo {address} {Firenze},\ \bibinfo {year} {1965})\ pp.\ \bibinfo {pages} {1--12}\BibitemShut {NoStop}%
\bibitem [{\citenamefont {Kerr}\ and\ \citenamefont {Schild}(2009)}]{Kerr:1965vyg}%
  \BibitemOpen
  \bibfield  {author} {\bibinfo {author} {\bibfnamefont {R.~P.}\ \bibnamefont {Kerr}}\ and\ \bibinfo {author} {\bibfnamefont {A.}~\bibnamefont {Schild}},\ }\href {https://doi.org/10.1007/s10714-009-0857-z} {\bibfield  {journal} {\bibinfo  {journal} {Gen. Rel. Grav.}\ }\textbf {\bibinfo {volume} {41}},\ \bibinfo {pages} {2485} (\bibinfo {year} {2009})}\BibitemShut {NoStop}%
\bibitem [{\citenamefont {Kerr}(1963)}]{Kerr:1963ud}%
  \BibitemOpen
  \bibfield  {author} {\bibinfo {author} {\bibfnamefont {R.~P.}\ \bibnamefont {Kerr}},\ }\href {https://doi.org/10.1103/PhysRevLett.11.237} {\bibfield  {journal} {\bibinfo  {journal} {Phys. Rev. Lett.}\ }\textbf {\bibinfo {volume} {11}},\ \bibinfo {pages} {237} (\bibinfo {year} {1963})}\BibitemShut {NoStop}%
\bibitem [{\citenamefont {Ay{\'o}n-Beato}\ \emph {et~al.}(2016)\citenamefont {Ay{\'o}n-Beato}, \citenamefont {Hassa{\"\i}ne},\ and\ \citenamefont {Higuita-Borja}}]{Ayon-Beato:2015nvz}%
  \BibitemOpen
  \bibfield  {author} {\bibinfo {author} {\bibfnamefont {E.}~\bibnamefont {Ay{\'o}n-Beato}}, \bibinfo {author} {\bibfnamefont {M.}~\bibnamefont {Hassa{\"\i}ne}},\ and\ \bibinfo {author} {\bibfnamefont {D.}~\bibnamefont {Higuita-Borja}},\ }\href {https://doi.org/10.1103/PhysRevD.94.064073} {\bibfield  {journal} {\bibinfo  {journal} {Phys. Rev. D}\ }\textbf {\bibinfo {volume} {94}},\ \bibinfo {pages} {064073} (\bibinfo {year} {2016})},\ \Eprint {https://arxiv.org/abs/1512.06870} {arXiv:1512.06870 [hep-th]} \BibitemShut {NoStop}%
\bibitem [{\citenamefont {Bonnor}(1969)}]{Bonnor:1969mfs}%
  \BibitemOpen
  \bibfield  {author} {\bibinfo {author} {\bibfnamefont {W.~B.}\ \bibnamefont {Bonnor}},\ }\href {https://doi.org/10.1007/BF01645484} {\bibfield  {journal} {\bibinfo  {journal} {Commun. Math. Phys.}\ }\textbf {\bibinfo {volume} {13}},\ \bibinfo {pages} {163} (\bibinfo {year} {1969})}\BibitemShut {NoStop}%
\bibitem [{\citenamefont {Aichelburg}\ and\ \citenamefont {Sexl}(1971)}]{Aichelburg:1970dh}%
  \BibitemOpen
  \bibfield  {author} {\bibinfo {author} {\bibfnamefont {P.~C.}\ \bibnamefont {Aichelburg}}\ and\ \bibinfo {author} {\bibfnamefont {R.~U.}\ \bibnamefont {Sexl}},\ }\href {https://doi.org/10.1007/BF00758149} {\bibfield  {journal} {\bibinfo  {journal} {Gen. Rel. Grav.}\ }\textbf {\bibinfo {volume} {2}},\ \bibinfo {pages} {303} (\bibinfo {year} {1971})}\BibitemShut {NoStop}%
\bibitem [{\citenamefont {Harada}(2025)}]{Harada:2025krm}%
  \BibitemOpen
  \bibfield  {author} {\bibinfo {author} {\bibfnamefont {J.}~\bibnamefont {Harada}},\ }\href {https://doi.org/10.1103/6ssh-b7lf} {\bibfield  {journal} {\bibinfo  {journal} {Phys. Rev. D}\ }\textbf {\bibinfo {volume} {112}},\ \bibinfo {pages} {024020} (\bibinfo {year} {2025})},\ \Eprint {https://arxiv.org/abs/2506.20878} {arXiv:2506.20878 [gr-qc]} \BibitemShut {NoStop}%
\bibitem [{\citenamefont {Taub}(1951)}]{Taub:1950ez}%
  \BibitemOpen
  \bibfield  {author} {\bibinfo {author} {\bibfnamefont {A.~H.}\ \bibnamefont {Taub}},\ }\href {https://doi.org/10.2307/1969567} {\bibfield  {journal} {\bibinfo  {journal} {Annals Math.}\ }\textbf {\bibinfo {volume} {53}},\ \bibinfo {pages} {472} (\bibinfo {year} {1951})}\BibitemShut {NoStop}%
\bibitem [{\citenamefont {Newman}\ \emph {et~al.}(1963)\citenamefont {Newman}, \citenamefont {Tamburino},\ and\ \citenamefont {Unti}}]{Newman:1963yy}%
  \BibitemOpen
  \bibfield  {author} {\bibinfo {author} {\bibfnamefont {E.}~\bibnamefont {Newman}}, \bibinfo {author} {\bibfnamefont {L.}~\bibnamefont {Tamburino}},\ and\ \bibinfo {author} {\bibfnamefont {T.}~\bibnamefont {Unti}},\ }\href {https://doi.org/10.1063/1.1704018} {\bibfield  {journal} {\bibinfo  {journal} {J. Math. Phys.}\ }\textbf {\bibinfo {volume} {4}},\ \bibinfo {pages} {915} (\bibinfo {year} {1963})}\BibitemShut {NoStop}%
\bibitem [{\citenamefont {Misner}(1963)}]{Misner:1963fr}%
  \BibitemOpen
  \bibfield  {author} {\bibinfo {author} {\bibfnamefont {C.~W.}\ \bibnamefont {Misner}},\ }\href {https://doi.org/10.1063/1.1704019} {\bibfield  {journal} {\bibinfo  {journal} {J. Math. Phys.}\ }\textbf {\bibinfo {volume} {4}},\ \bibinfo {pages} {924} (\bibinfo {year} {1963})}\BibitemShut {NoStop}%
\bibitem [{\citenamefont {Heusler}(1996)}]{Heusler:1996}%
  \BibitemOpen
  \bibfield  {author} {\bibinfo {author} {\bibfnamefont {M.}~\bibnamefont {Heusler}},\ }\href {https://doi.org/10.1017/CBO9780511661396} {\emph {\bibinfo {title} {{Black Hole Uniqueness Theorems}}}}\ (\bibinfo  {publisher} {{Cambridge Univ. Press}},\ \bibinfo {year} {1996})\BibitemShut {NoStop}%
\bibitem [{\citenamefont {Boyer}\ and\ \citenamefont {Lindquist}(1967)}]{Boyer:1966qh}%
  \BibitemOpen
  \bibfield  {author} {\bibinfo {author} {\bibfnamefont {R.~H.}\ \bibnamefont {Boyer}}\ and\ \bibinfo {author} {\bibfnamefont {R.~W.}\ \bibnamefont {Lindquist}},\ }\href {https://doi.org/10.1063/1.1705193} {\bibfield  {journal} {\bibinfo  {journal} {J. Math. Phys.}\ }\textbf {\bibinfo {volume} {8}},\ \bibinfo {pages} {265} (\bibinfo {year} {1967})}\BibitemShut {NoStop}%
\bibitem [{\citenamefont {Astefanesei}\ \emph {et~al.}(2005)\citenamefont {Astefanesei}, \citenamefont {Mann},\ and\ \citenamefont {Radu}}]{Astefanesei:2004kn}%
  \BibitemOpen
  \bibfield  {author} {\bibinfo {author} {\bibfnamefont {D.}~\bibnamefont {Astefanesei}}, \bibinfo {author} {\bibfnamefont {R.~B.}\ \bibnamefont {Mann}},\ and\ \bibinfo {author} {\bibfnamefont {E.}~\bibnamefont {Radu}},\ }\href {https://doi.org/10.1088/1126-6708/2005/01/049} {\bibfield  {journal} {\bibinfo  {journal} {JHEP}\ }\textbf {\bibinfo {volume} {01}},\ \bibinfo {pages} {049}},\ \Eprint {https://arxiv.org/abs/hep-th/0407110} {arXiv:hep-th/0407110} \BibitemShut {NoStop}%
\bibitem [{\citenamefont {Myers}\ and\ \citenamefont {Perry}(1986)}]{Myers:1986un}%
  \BibitemOpen
  \bibfield  {author} {\bibinfo {author} {\bibfnamefont {R.~C.}\ \bibnamefont {Myers}}\ and\ \bibinfo {author} {\bibfnamefont {M.~J.}\ \bibnamefont {Perry}},\ }\href {https://doi.org/10.1016/0003-4916(86)90186-7} {\bibfield  {journal} {\bibinfo  {journal} {Annals Phys.}\ }\textbf {\bibinfo {volume} {172}},\ \bibinfo {pages} {304} (\bibinfo {year} {1986})}\BibitemShut {NoStop}%
\end{thebibliography}
\end{document}